\documentclass{aastex}
\usepackage{psfig}   
\usepackage{subfigure}
\usepackage{spr-astr-addons}
\usepackage{url}

\RequirePackage{color}

\begin{document}

\title{  Evolution of superhigh magnetic fields of
magnetars }
\slugcomment{Not to appear in Nonlearned J., 45.}
\shorttitle{The electron capture rate onto  protons }
\shortauthors{Z. F. Gao et al.}

\author{Z. F. Gao\altaffilmark{1,2,3}}
\altaffiltext{1}{Urumqi Observatory, NAOC, 40-5 South Beijing Road,
Urumqi Xinjiang, 830011, China zhifu$_{-}$gao@uao.ac.cn}
\altaffiltext{2}{Graduate University of the Chinese
   Academy of Scienes, 19A Yuquan road, Beijing, 100049, China}
\altaffiltext{3}{Department of Astronomy, Nanjing University, Nanjing, 210093, China}
\author{ N. Wang \altaffilmark{1}}
\affil{ Urumqi Observatory, NAOC, 40-5 South Beijing Road, Urumqi
Xinjiang, 830011, China}
\author{ J. P. Yuan \altaffilmark{1}}
\affil{Urumqi Observatory, NAOC, 40-5 South Beijing Road, Urumqi
Xinjiang, 830011, China}
\author{ L. Jiang \altaffilmark{3}}
\affil{ XinJiang Education Institute, 333 Guang Ming Road, Urumqi
Xinjiang, 830043, China }
\author{ D. L. Song \altaffilmark{4}}
\affil{ The Information Engineering University, 63 Science Road,
ZhengZhou Henan, 450001, China }
\author{E. L. Qiao \altaffilmark{5}}
 \affil {Yunnan Observatory, Kunming, Yunnan, 650000, China}
\begin{abstract}
In this paper, we consider the effect of Landau levels on the
decay of superhigh magnetic fields of magnetars.  Applying
${}^3P_2$ anisotropic neutron superfluid theory yield a
second-order differential equation for a superhigh magnetic field
$B$ and its evolutionary timescale $t$.  The superhigh magnetic
fields may evolve on timescales $\sim (10^{6}-10^{7})$ yrs for
common magnetars.  According to our model, the activity of a
magnetar may originate from instability caused by the high electron
Fermi energy.

\end{abstract}


\section{Introduction}
It is now universally accepted that pulsars are neutron
stars (NSs) with extremely strong magnetic fields.  The
surface magnetic field manifests itself as synchrotron
radiation from the pulsar magnetosphere, where the field
magnitude for young NSs has been estimated to be $(10^{12}
\sim 10^{13})$ G. The internal magnetic field is expected
to be even higher \citep{mih90}.  The magnetic field is the
main energy source of all the persistent and bursting
emission observed in anomalous X-ray pulsars(AXPs) and soft
gamma-ray repeaters (SGRs)\citep{dun92, pac92, mer08}.  AXPs
and SGRs are considered to be Magnetar candidates, namely
isolated neutron stars powered by the decay of huge ($B >
B_{cr}$ = 4.414 $\times 10^{13}$ G) magnetic fields inferred
from the rotation under the so-called `standard assumption':
$\alpha- \Omega$ dynamo \citep{dun92, tho95, tho96}.  Their
common properties are: stochastic outbursts (lasting from
days to years) during which they emit very short X/$\gamma$-
ray bursts; rotational periods in a narrow range $P
\sim$(6- 12) s; compared to other isolated neutron stars, large
period derivatives of $(10^{-13} \sim 10^{-10})$ s~s$^{-1}$;
rather soft (at least for AXPs) X-ray spectra that can be
fitted by the sum of a black body ($kT \sim$ 0.5 keV) and a
power law component (power law index $\Gamma \sim $2- 4), and,
in some cases, associated with supernova remnants (SNRs)
\citep{dun92, mer08}.

Up to now (3 October 2010), there are nine SGRs (seven
conformed)and twelve AXPs (nine conformed) detected, so
statistically investigating their persistent parameters is
available.  Observationally, all known magnetars are X-ray
pulsars with luminosities of $L_{\rm X}\sim (10^{32}\sim 10^{36})$
erg~s$^{-1}$ \citep{rea10}, their high luminosities together
with the lack of evidence for accretion from a stellar companion,
result in the conclusion that the energy reservoir fueling the
magnetar activity is their superhigh magnetic fields ($10^{14}
\sim 10^{15}$) G \citep{dib08, tho96, mer08, rea10}.  Such a huge
energy reservoir ($E \sim 3 \times 10^{47}(B/10^{15}{\rm G})^{2}(R/10{\rm km})
^{3}$ erg) is sufficient to power not only the persistent surface
X-ray emission but also a steady stream of low-amplitude Alfven
waves into the magnetosphere for $\sim 10^{4}$ yrs \citep{tho95, tho96}.

In order to provide a detailed view of the actual evolution of a
magnetar, we should consider the effective electron capture rate
$\Gamma_{\rm eff}$ (the effective number of electrons captured by one
proton per second) due to the existence of the Landau levels of
charged particles \citep{gao10a}.  However, only the electrons
occupying large Landau levels with high energy ($E > Q$, $Q$ is
the threshold of energy) are allowed to participate in the process
$e^{-} + p \rightarrow n+ \nu_{\rm e}$ \citep{yak01}.  In other words,
whether an electron could be captured depends not only on the
electron's energy $E_{\rm e}$ but also on the number of the Landau
level it occupies.  Based on the assumption of the Landau level
effect coefficient $q $, we define the effective electron capture
rate $\Gamma_{\rm eff}$ as: $\Gamma_{\rm eff}= q \Gamma$ \citep{gao10a}.
With respect to the electron Fermi energy $E_{\rm F}(\rm e)$, our point
of view is that in the case of an intense magnetic field, the
stronger the magnetic field, the higher the electron Fermi energy
released from the magnetic field energy \citep{gao10b}.

The remainder of this paper is organized as follows: in $\S$ 2 a
review on the evolution of superhigh magnetic fields of magnetars
is given; in $\S$ 3 we investigate the relation between $E_{\rm F}(\rm e)$
and $B$, and gain a  schematic diagram of $L_{X}$ as a function of
$B$; in $\S$ 4 a second-order differential equation for a superhigh
field and its evolutionary timescale is given; discussions and
conclusions are presented in $\S$ 5; in Appendix $A$ we consider the
effect of a superhigh magnetic field on the proton fraction $Y_{\rm p}$
and gain a concise formula  of $E_{\rm F}(\rm e)$, and in Appendix $B$ we
give an explanation for why a wrong notion that $E_{\rm F}(\rm e)$
decreases with increasing $B$ is used as yet.

\section{The evolution of magnetic fields inside a magnetar}

The evolution of magnetic fields of a magnetar is a rather
challenging subject, with strong observational clues and
involving complex physics.  The strong evidences for spontaneous
evolution of magnetic fields inside magnetars are mainly that:
SGRs and AXPs appear to radiate substantially more power than
available from their rotational energy loss; the dipole field
inferred from the observed torque is stronger than that of all
other NSs, $\sim 10^{14}-10^{15}$ G, which make it nature to
accept the argument of Thompson $\&$ Duncan (1992) that these
sources are in fact `magnetars', powered by the dissipation of
their magnetic energy; and torque changes of both signs,
associated with outbursts from these sources, etc. In order to
explain the evolution of superhigh magnetic fields inside
magnetars, different models have been proposed recently, as
partly listed below.

In the twisting magnetic field model, the magnetic field is
supposed to be dominated by a toroidal component larger than
the external dipole.  The persistent quiescent X-ray emission
of a magnetars could be brought out by the twisting of the
external magnetic field caused by the motions of the star
interior.  The twisting crustal motions generally drive
currents outside a magnetar, and generate X-rays \citep{tho00}.
 Furthermore, a discussion about the globally-twisted magnetosphere
was proposed by Thompson et al (2002).  However, up to now, a
quantitative calculation of X-ray luminosity in the twisting
magnetic field model is still up in the air.

In the thermal evolution model, the main points of view include:
The field could decay directly as a result of the non-zero
resistivity of the matter Ohmic decay or ambipolar diffusion, or
indirectly as a result of Hall drift producing a cascade of the
field to high wave number components, which decay quickly through
Ohmic decay \citep{gol92, pon07, rhe03}; magnetic field decay can
be a main source of internal heating; the enhanced thermal
conductivity in the strongly magnetized envelope contributes
to raise the surface temperature \citep{hey97, hey98}, etc.

In currently existing magnetar models, the evolution timescale
of superhigh magnetic fields of magnetars is estimated just
according to conservation of energy, rather than depending on
an accurate equation of $B$ and its evolution timescale t.
For example, a strong dependence of neutron star surface
temperature on the dipolar component of the magnetic
field $B_{\rm dip}$ for stars with $B_{\rm dip}> 10^{13}$ G reveals
that the thermal evolution is almost fully controlled by the
amount of magnetic field energy the star has stored in its
crust by the time the star has reached an age of $\sim 10^{5}$
yrs (earlier, for magnetars) \citep{pon07}. One more example,
while the field decay is governed by Ohmic dissipation and Hall
drift, the most interrelated process inside a magnetar is
ambipolar diffusion, which has a characteristic timescale
\begin{equation}
t_{\rm amb}\sim 10^{4}\times(B_{\rm core}/10^{15}{\rm G})^{-2} \rm yrs,
\end{equation}
where $B_{\rm core}$ denotes the core magnetic field \citep{tho96}.
In this article, unlike the previous  magnetar model, we calculate
the decay timescale of superhigh magnetic fields of magnetars
by employing the Landau level effect coefficient $q $ and the
effective electron capture rate $\Gamma_{\rm eff}$ \citep{gao10a}.
For further details, see $\S$ 4.

With respect to the origin of superhigh magnetic fields of
magnetars, a more popular hypothesis is that magnetars are
formed from rapidly rotating proto-neutron stars, differential
rotation and convection would result in an efficient
$\alpha-\Omega$ dynamo \citep{dun92, dun96, tho93, vin06}. The
dynamo responsible for the superhigh magnetic field generation
requires that magnetars be born with ultrashort initial spin
periods $P_{\rm i}\sim $1 ms, the associated free energy is $E_{\Omega i}
\sim 10^{52}(P_{\rm i}/1{\rm ms})^{-2}$ erg, and the magnified field is
as strong as $\sim 3 \times 10^{17}(P_{\rm i}/1 ms)^{-1}$ G \citep{tho95,
tho96}.  However, there is as yet no mechanism to explain such a
high efficiency of energy transformation. When investigating solar
flare using the $\alpha-\Omega$ model, apart from the difficulty
in explaining such a high energy transformation efficiency, there
are also many other problems to be settled at present.  Note, up
to now, there is no observational indication on the existence of
fields $B \geq 10^{16}$ G in the interior of a NS.  Furthermore,
the explosion energies of these supernova remnants associated
with AXPs and SGRs (AXP 1E 1841-045, AXP 1E 2259+586, and SGR
0526-26), are near to the canonical supernova explosion energies
of $\sim 10^{51}$ erg, suggesting $P_{\rm i}\geq$ 5 ms \citep{vin06}.
Enough has been said to show that $\alpha-\Omega$ dynamo is a
mere assumption, laking observational support.

Although the origin of superhigh magnetic fields of
magnetars, as mentioned above, is uncertain, a possible
mechanism that superhigh magnetic fields of magnetars
originate from the induced magnetic fields at a moderate
lower temperature, has been proposed \citep{pen07, pen09}.
It is universally supposed that ${}^3P_2$ anisotropic neutron
superfluid could exist in the interior of a NS, where the mean
thermal kinetic energy of nucleons is far less than nucleon
interaction energy (in the order of $\sim$ 1 MeV) due to
strong nucleon-nucleon attractive interaction (the mean
distance of nucleons is $\sim $1 fm). It is this strong nucleon-
nucleon interaction of attraction that results in the formations
of proton Cooper pairs and ${}^3P_2$ anisotropic neutron Cooper
pairs. The nucleonic ${}^3P_2$ pairing gaps in NSs have been
investigated by Elgar${\o}$y et al (1996) in detail.  Their
important results may be briefly summarized as follows:

$(1).$ The ${}^3P_2$ neutron pair energy gap $\Delta({}^3P_2)$
appears in the region 1.32 $< k_{\rm F}$ (fm$^{-1}$)$<$ 2.34, where
$\Delta({}^3P_2)$ is an increasing function of the density for
the region 1.32 $< k_{\rm F}$(fm$^{-1}$) $<$ 1.8 and it is a rapidly
decreasing function of the density for the region 2.1 $< k_{F}$
(fm$^{-1}$)$<$ 2.34, where $k_{\rm F}$ is the Fermi wave number of the
neutrons.

$(2).$	The maximum of the ${}^3P_2$ neutron pair energy gap
$\Delta_{\rm max}({}^3P_2)$is about 0.048 MeV at $k_{\rm F}$= 1.96 fm$^{-1}$.

$(3).$  $\Delta({}^3P_2)$ is almost a constant about the maximum
with error less than $3\%$ in the rather wide range 1.8 $ < k_{\rm F}$
(fm$^{-1}$)$<$ 2.1, corresponding to the density region 3.3 $\times
10^{14} < \rho$ (g~cm$^{-3}) < $5.2 $\times 10^{14}$.

Based on the analysis above, the critical temperature of the
${}^3P_2$ neutron Cooper pairs can be evaluated as follows:$T_{cn}=
\Delta_{\rm max}({}^3P_2)/2k \approx$ 2.78$\times 10^{8}$K,
the maximum of internal temperature $T$ (rather than the
maximum of core temperature) can not exceed $T_{cn}$
according to our model \citep{pen06, pen09}.  A NS is
often approximately considered as a system of magnetic
dipoles because of the existence of ${}^3P_2$ anisotropic
neutron superfluid in its interior.  In the presence of a
background magnetic field, the dipoles have a tendency to
be aligned along the direction of this field.  If temperature
falls below a critical temperature, a phase transition from
paramagnetism to ferromagnetism could occur, which provides
a possible explanation for superhigh magnetic fields of
magnetars.  The maximum induced magnetic field is estimated
to be (3.0 $\sim$ 4.0)$)\times 10^{15}$ G \citep{pen06}.

In the case of field-free, for the direct Urca reactions to
take place, there exists the following inequality among the
Fermi momenta of the proton ($P_{\rm F}$), the electron ($K_{\rm F}$)
and the neutron ($Q_{F}$): $P_{\rm F}+ K_{\rm F}\geq Q_{\rm F}$ required
by momentum conservation near the Fermi surface.  Together with
the charge neutrality condition, the above inequality brings
about the threshold for proton concentration $Y_{\rm p}= n_{\rm p}/
(n_{\rm p}+ n_{\rm n})\geq \frac{1}{9}$ = 0.11, this means that, in the
field-free case, direct Urca reactions are strongly suppressed
in the degenerate $n-p-e$ system (Baiko \& Yakovlev 1999; Lai \&
 Shapiro 1991; Leinson \& P$\acute{e}$rez. 1997; Yakovlev
et al 2001).  However, when in a superhigh magnetic field $B \gg
B_{\rm cr}$, things could be quite different. A superhigh magnetic
field can generate a noticeable magnetic broadening of the
direct Urca process threshold, as well as of the thresholds
of other reactions in the interior of a magnetar \citep{yak01}.
If the magnetic field is large enough for protons and electrons
to be confined to the lowest landau levels ($n$=0, 1), the
field-free threshold condition on proton concentration no
longer holds, and direct Urca reactions are open for an
arbitrary proton concentration (Leinson \& P$\acute{e}$rez.
1997).  Strong magnetic field can alter matter compositions
and increase phase pace for protons which leads to the
increase of $Y_{\rm e}$ \citep{lai91}.  By strongly modifying the
phase spaces of protons and electrons, ultrastrong magnetic
fields ($\sim 10^{20}$ G) can cause a substantial $n \rightarrow
p$ conversion, as a result, the system is converted to highly
proton-rich matter with distinctively softer equation of states,
compared to the field-free case\citep{cha97}.  Though magnetic
fields of such magnitude inside NSs are unauthentic, and are not
consistent with our model ($B\sim 10^{14-15}$ G), their calculations
are useful in supporting our assumption: Direct Urca reactions
are expected to occur inside a magnetar.  According to our model,
as soon as the energy of electrons near the Fermi surface are
higher than the Fermi energy of neutrons ($E_{\rm F}(\rm n)\approx$
60 MeV see Chapter 11 of Shapiro \& Teukolsky 1983), the
process $e^{-}+ p \rightarrow n+ \nu_{\rm e}$ will dominate.  By
colliding with the neutrons produced in the process of $n +
(n\uparrow n\uparrow)\longrightarrow n+ n+ n$, the kinetic
energy of the outgoing neutrons will be transformed into
thermal energy and then transformed into the radiation energy
with X-ray and soft $\gamma$-ray \citep{pen09}.  The ${}^3P_2$
Cooper pairs will be destroyed quickly by these high-energy
outgoing neutrons.  Thus the anisotropic superfluid and the
superhigh magnetic fields induced by the ${}^3P_2$ Cooper
pairs will disappear.

\section{The electron Fermi energy and direct Urca process}
This section is composed of two subsections.  For each
subsection we present different methods and considerations.
\subsection{The Fermi energy of electrons in superhigh
magnetic fields}
As shown in Fig.1, in the presence of an extremely strong
magnetic field, the electrons are situated in disparate energy
states in order one by one from the lowest energy state up to
the Fermi energy (the highest energy) with the highest momentum
$p_{\rm F}(z)$ along the magnetic field, according to the Pauli
exclusion principle.

\begin{figure}[th]
\centering
 \vspace{0.5cm}
 \includegraphics[width=0.98\columnwidth,angle=0]
 {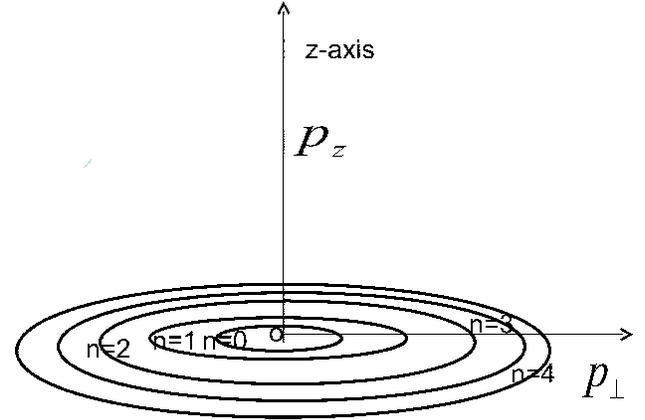}
 \caption{The schematic diagram of Landau quantization. The
applied magnetic field is along the $z$-axis direction, the energy
levels perpendicular to the direction of applied magnetic field
are quantized.  The radius of the cross section is $p_{\perp}$,
$p_{z}$ is the momentum along $z$-axis direction.}
 \label{fig:landau}
 \end{figure}

In the interior of a magnetar, different forms of intense magnetic
field could exist simultaneously and a high $E_{\rm F}(\rm e)$ could be
generated by the release of magnetic field energy.  When $B \gg
B_{\rm cr}$, the Landau column becomes a very long and very narrow
cylinder along the magnetic field, the electron Fermi energy is
determined by
\begin{eqnarray}
&&\frac{3\pi}{B^{*}}(\frac{m_{\rm e}c}{h})^{3}(\gamma_{\rm e})^{4}\int_{0}^{1}
(1-\frac{1}{\gamma^{2}_{\rm e}}-\chi ^{2})^{\frac{3}{2}}d\chi\nonumber\\
&&-2\pi\gamma_{\rm e}(\frac{m_{\rm e}c}{h})^{3}\sqrt{2B{*}}= N_{A}\rho Y_{\rm e},
\end{eqnarray}
where $B^{*}$, $\chi$ and $\gamma_{\rm e}$ are three non-dimensional
variables, which are defined as ${B^{*}= B/B_{\rm cr}}$, $\chi =(\frac{p_
{z}}{m_{\rm e}c})/(\frac{E_{\rm F}}{m_{\rm e}c^{2}})= p_{z}c/E_{\rm F}$ and $\gamma_{\rm e}
= E_{\rm F}/m_{\rm e}c^{2}$, respectively; $1/\gamma^{2}_{\rm e}$ is  the
modification factor; $N_{A}=$ 6.02 $\times 10^{23}$ is the Avogadro
constant; $Y_{\rm e}=Y_{\rm p}= Z/A$, here $Y_{\rm e}$, $Y_{\rm p}$, $Z$ and $A$ are
the electron fraction, the proton fraction, the proton number and
nucleon number of a given nucleus, respectively \citep{gao10b}.  From
Eq.(2), we gain the relations of $E_{\rm F}(\rm e)$ vs. $B$ and $E_{\rm F}(\rm e)$
vs. $\rho$, which are shown in Fig. 2.

\begin{figure}[th]
\centering
 \vspace{0.5cm}
\subfigure[]{
    \label{evsb:a} 
    \includegraphics[width=7.2cm]{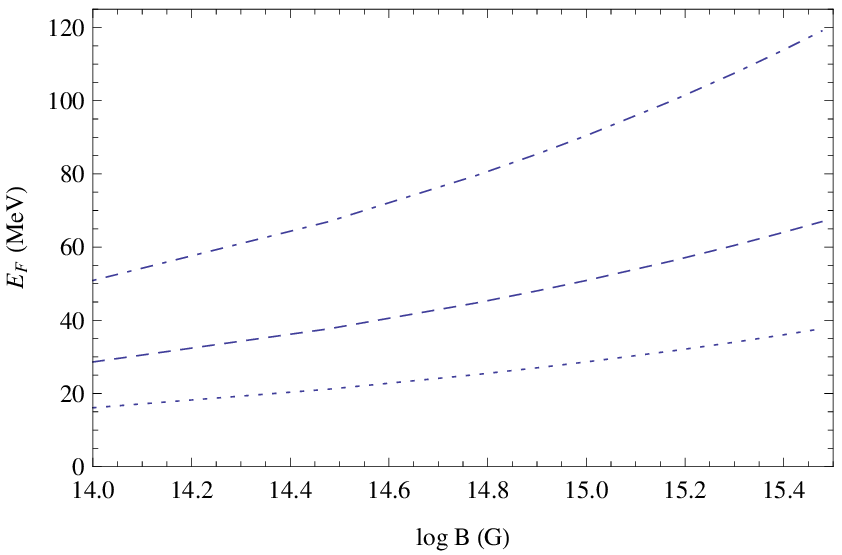}}
  \hspace{0.2mm}
  \subfigure[]{
    \label{evsb:b} 
    \includegraphics[width=7.0cm]{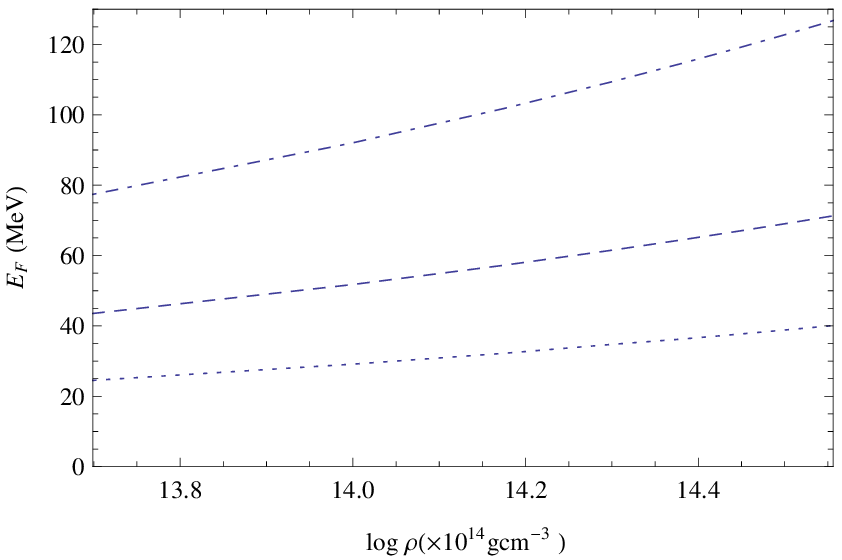}}
 \caption{Top, the relation of $E_{\rm F}(\rm e)$ and $B$. The range of
$B$ is (1.0$\times 10^{14}\sim$ 3.0$\times 10^{15}$)G, $\rho$
=2.8 $\times 10^{14}$ g~cm$^{-3}$, dot-dashed line, dashed
line and dotted line are for $Y_{\rm e}$= 0.1, $Y_{\rm e}$= 0.01 and
$Y_{\rm e}$= 0.001, respectively.  Bottom, the relation of $E_{\rm F}(\rm e)$
and $\rho$. The range of $\rho$ is (0.5$\sim$ 3.6)$\times
10^{14}$ g~cm$^{-3}$, $B$= 3.0 $\times 10^{15}$ G, dot-dashed
line, dashed line and dotted line are for $Y_{\rm e}$= 0.1, $Y_{\rm e}$
=0.01 and $Y_{\rm e}$= 0.001, respectively.}
 \label{fig:Fermi}
 \end{figure}

Seeing from Fig.2, it's obvious that $E_{\rm F}(\rm e)$ increases
with the increase in $B$ when $\rho$ and $Y_{\rm e}$ are given;
$E_{\rm F}(\rm e)$ also increases with the increasing $\rho$ when
$B$ and $Y_{\rm e}$ are given. The high Fermi energy of electrons
could be from the release of the magnetic energy.  A possible
interpretation of high $E_{\rm F}(\rm e)$ is given as follows:  an
envelope of the Landau circles with maximum quantum number
$n_{\rm max}$ will approximately form a sphere, i.e. Fermi sphere.
For a given electron number density with a highly degenerate
state in the interior of a NS, the stronger the magnetic field,
the larger the maximum of $p_{z}$ is, hence the lower the
number of states in the $x-y$ plane according to the Pauli
exclusion principle.  In other words, $n_{max}$ and the number
of electrons in the $x-y$ plane decrease with the increase of
$B$, the radius of the Fermi sphere $p_{\rm F}$ is expanded which
implies that the electron Fermi energy $E_{\rm F}(\rm e)$ also
increases.  The higher the Fermi energy $E_{\rm F}(\rm e)$, the more
obvious the `expansion' of the Fermi sphere is, however, the
majority of the momentum space in the Fermi sphere is empty for
not being occupied by electrons.  However, one incorrect notion
that $E_{\rm F}(\rm e)$ decreases with increasing $B$ in an intense
field ($B\gg B_{\rm cr}$) has been universally adopted for a long
time. With respect to this incorrect notion, further details
are presented in Appendix $B$.

\subsection{Direct Urca process in superhigh magnetic fields}
We assume that at zero-temperature the NS is $\beta$-stable,
but at non-zero temperature ($kT \ll E_{\rm F}(\rm i), \rm i= n, p, e$,
$k =$1.38 $\times 10^{-16}$ erg~ K$^{-1}$ is the Boltzmann
constant), reactions $e^{-}+ p \rightarrow n+ \nu_e$ and
$n\rightarrow e^{-}+ p + \nu^{-}_{e}$ proceed near the Fermi
energies $E_{\rm F}(\rm i)$ of the participating particles.  The
electron capture rates can be calculated as follows:
\begin{equation}
d\Gamma= \Lambda I= \frac{2\pi}{\hbar}\frac{G_{F}^{2}C_{V}^{2}
(1+ 3a^{2})}{(2\pi^{2}\hbar^{3}c^{3})^{2}}I
\end{equation}
\begin{eqnarray}
&&I= \int_{60}^{E_{\rm F}(\rm e)}(E_{\rm e}^{2}-m^{2}_{\rm e}
c^{4})^{\frac{1}{2}}E_{\rm e}\nonumber\\
&&(E_{\rm e}-60)^{2}\frac{1}{e^{\frac{E{\rm e}-E_{\rm F}(\rm e)}
{kT}}+1}\frac{1}{e^{\frac{Q-E_{\rm e}}{kT}}+1}dE_{\rm e},
\end{eqnarray}
where $Q = E_{\rm F}(\rm n)-E_{\rm F}(\rm p)\simeq $ 60 MeV, $\Lambda \approx$
0.018 (MeV)$^{-5}$ s$^{-1}$, $E_{\rm F}(\rm e)$ =40$(B/B_{\rm cr})
^{\frac{1}{4}}$ MeV (c.f. Gao et al 2010a and Appendix
$A$ of this paper), and other terms appearing in Eqs.(3-4)
have already been defined in Chapter 18 of \citep{sha83}.
In this paper, for the purpose of convenient calculation,
we set $\rho= \rho_{0}$.  The range of $B$ is assumed to be
(2.2346 $\times 10^{14}\sim$ 3.0 $\times 10^{15}$) G corresponding
to $E_{\rm F}(e)\sim$(60 $\sim$114.85) MeV, where 2.2346 $\times
10^{14}$ G is the minimum of $B$ denoted as $B_{\rm f}$.  When $B$
drops below $B_{\rm f}$, the direct Urca process is quenched
everywhere in the magnetar interior\citep{gao10a}.  Combining
the relation
\begin{equation}
\langle E_\nu \rangle= \int_{Q}^{E_{\rm F}(\rm e)}S(E_{\rm e}-Q)^{3}
E_{\rm e}(E_{\rm e}^{2}-m_{\rm e}^{2}c^{4})^{\frac{1}{2}}dE_{\rm e}/I,
\end{equation}
with the relation
\begin{equation}
\langle E_{\rm n}\rangle = E_{\rm F}(\rm e)- \langle E_{\nu}\rangle- 1.29 MeV,
\end{equation}
gives the mean kinetic energy of neutrons.  The average X-ray
luminosity $L_{\rm X}$ can be expressed as follows:
\begin{equation}
 L_{\rm X}= \Gamma_{\rm eff} n_{\rm p} V({}^3P_2) \langle E_{\rm n} \rangle
 = q \Gamma n_{\rm p} V({}^3P_2) \langle E_{\rm n} \rangle,
 \end{equation}
where $q$ is the Landau level effect coefficient, $V({}^3P_2)$
denotes the volume of ${}^3P_2$ anisotropic neutron superfluid
($V({}^3P_2)= \frac{4}{3}\pi R_{5}^{3}, R_{5}= R/10^{5}\rm cm $ )
and $n_{\rm p}= n_{\rm e}$= 9.6 $\times 10^{35}$ cm$^{-3}$.  For a
magnetar with initial magnetic field $B_{0}$= 3.0 $\times
10^{15}$ G, if its initial X-ray luminosity $L_{\rm X0}$ is assumed
to be (1.0$\sim$9.0)$\times 10^{36}$ erg~ s$^{-1}$, then $q$ is
estimated to be (0.216 $\sim$ 1.94)$\times 10^{-18}$ and the
initial effective electron capture rate is (0.221 $\sim$ 1.98)$
\times 10^{-11}$ s$^{-1}$.  We can also construct a schematic
diagram of the average X-ray luminosity as a function of
magnetic field for the process of electron capture, which is
shown as in Fig.4.

 \begin{figure}[th]
\centering
  \vspace{0.5cm}
  \includegraphics[width=6.0cm,angle=-90]{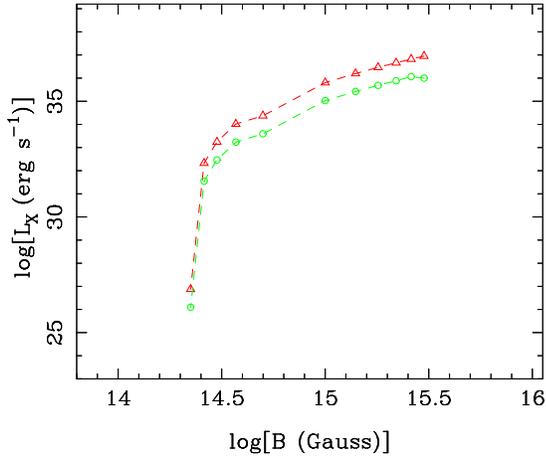}
  \caption{The schematic diagram of $L_{\rm X}$ vs. $B$.  Triangles
and circles mark the values of variables corresponding to $q$
=1.94 $\times 10^{-18}$ and 2.16 $\times 10^{-19}$, respectively.
The ranges of $B$ is assumed to be (2.24 $\times 10^{14}\sim
$ 3.0 $\times 10^{15}$ Gauss) considering that, when $B \leq B_{\rm f}$,
 the direct Urca process ceases, while the modified Urca process
 still occurs, from which weaker X-ray and weaker neutrino
 flux are produced.}
  \label{ Lx-B. fig}
\end{figure}

As discussed above, the decay of magnetic fields is
the ultimate energy source of all the persistent and bursting
emission observed in AXPs and SGRs.  In Table 1, the related
persistent parameters of the conformed magnetars with observed
X-ray fluxes are listed according to observations performed
in the last two decades.

\begin{table*}[th]
\small
\caption{AXP/SGR persistent parameters. \label{tb3}}
\begin{tabular}{@{}crrrrr@{}}
\tableline
Name&  $\dot{P}$ &$T_{\rm BB}$&   $B $&$L_{\rm X}$\\
\tableline
SGR1627-41 & 1.9(4) &  $*$ &2.25& 0.025\\
SGR1806-20 &54.9& 7.54&21 &(50$^{a}$)\\
SGR1900+14& 7.783 &4.98&6.42& 1.8-2.8\\
SGR0526-66& 6.5& 6.14 & 7.32& 2.1\\
CXOUJ0100&1.88(8) & 4.41& 3.94& 0.78 \\
4U 0142+61 &0.196 &4.60 & 1.32& $>$0.53\\
1E 1841-045&4.1551 &5.10 & 7.08& 2.2 \\
1E 2259+586 & 0.048&  $*$ &0.59& 0.18\\
1RXS J1708  &1.945 &5.29 & 4.68& 1.9 \\
\tableline
CXO$^{t}$J1647 &0.24& 0.63& 1.61& 0.26 \\
1E$^{t}$1547.0-5408 &2.318 &0.43 & 2.21& 0.031 \\
XTE$^{t}$J1810-197 &0.81 &0.68& 1.7& 0.031\\
\tableline
\end{tabular}
\tablecomments{The units of $\dot{P}$, $T_{\rm BB}$, $B$ and $L_{\rm X}$ are $10^{-11}$ s~s$^{-1}$,\\
$10^{6}$ K, $10^{14}$ G and $10^{35}$ erg~s$^{-1}$, respectively, where $T_{\rm BB}$\\
is the surface temperature of a magnetar.  The sign $a$ de\\
-notes: from Thompson \& Duncan 1996.  The sign $t$ de\\
-notes: transient AXP.  All data are from the McGill AXP\\
/SGR online catalog of 3 October 2010 (http: www.physics.\\
mcgill. ca/$^{\sim}$pulsar/magnetar/main.html) except for $L_{\rm X}$ of\\
SGR 1806-20.  The data of column 3 are gained from the \\
original data by using 1 KeV $\simeq$ 1.16 $\times 10^{7}$ K }
\end{table*}
From Table 1, we see weak correlation between $T_{BB}$ and
$B$.  Magnetars are massive cooling neutron stars, according
to neutron star cooling theory (modified Urca reactions), the
typical magnetar internal temperature is about 3 $\times 10^{8}$
K, its surface temperature is lower than its internal temperature
by two orders of magnitude \citep{yak01}, which is consistence
with the surface temperature data in Table 1.  The assumption
that ${}^3P_2$ anisotropic neutron superfluid could exist in the
interior of a NS is supported by the data of column 3 in Table 1.

The inverse $\beta$-decay and related reactions in superhigh
magnetic fields have been studied since late 1960's
\citep{bai99, ban98, can71, dor85, lai91, lat91}. The
conventional wisdom is that the neutrino emissivity $Q_{\nu}
\propto T_{9}^{6}$ ($kT_{9}\sim$ 0.086 MeV) in the direct Urca
process, and the `standard magnitude' of $Q_{\nu}$ is $\sim
10^{27}T_{9}^{6}$ erg~cm$^{-3}$~s$^{-1}\sim 10^{45}T_{9}^{6}$
erg~s$^{-1}$, where $T_{9}$ is the temperature in units of
$10^{9}$ K and it is assumed that the superhigh magnetic field
permeates the entire volume ($R_{star}\sim 10^{6}$ cm) of the
magnetar interior.  With respect to the `standard magnitude' of
$Q_{\nu}$, the main assumptions include: $E_{\rm i}\sim E_{\rm F}(\rm i)$
(i=e, p, n, $\nu$); $F_{\rm F}(\rm e)$ decreases with increasing
$B$; $E_{\rm F}(p)\sim E_{\rm F}(\rm n) \sim E_{\rm F}(\nu)\sim kT_{9}$.  From
the expression $Q_{\nu}\propto T_{9}^{6}$, it is easy to imagine
that the neutrino flux comes from the thermal energy in the
magnetar interior, rather than from the free energy of the
superhigh magnetic field.  Why should we say this?  Results
of our analysis appear below: In the interior of a NS, the
majority of fermions are situated in the bottom of Fermi sea
\citep{lan65}, their energies are far less than their Fermi
energies, only in this case can we write $E_{\rm i}\sim kT$; however,
in the vicinity of Fermi surface, there are but few fermions
with energies $E_{\rm i}\sim E_{\rm F}(\rm i)$.  We must bear in mind that
the minimum of $E_{n}$ of the outgoing neutrons in the process
of electron capture is no less than the neutron Fermi energy
($E_{\rm F}(\rm n)\geq $ 60 MeV \citep{sha83}), otherwise the outgoing
neutrons can not escape from the surface of Fermi sea, thus the
inverse $\beta$-decay will not occur.  In the interior of a
magnetar where the direct Urca reaction takes place, neutrons
can be sorted into two kinds according to their energies: low-
energy neutrons with thermal kinetic energy $\sim kT$ and high-
energy outgoing neutrons with energy $\sim E_{\rm F}(\rm n)$, for
neutrinos, the same are true.  Although the energies of
outgoing neutrons and outgoing neutrinos are directly from
the energies of electrons participating in direct Urca
reaction, $L_{\rm X}$ and $Q_{\nu}$ are ultimately determined by
$B$, rather than by $T$, the reason is that magnetic field
energy is the main source of emissions, whereas $T$ is only
equivalent to background temperature, therefore both $L_{\rm X}$
and $Q_{\nu}$ are just weak functions of $T$.  With regard to
the calculations of X-ray luminosity and neutrino luminosity
in superhigh magnetic fields, for further details, refer to
Gao et al (2010a).  The magnetic field distribution of a
magnetar, though attracting general attentions of researchers,
still lacks a definite conclusion now.  If the direct Urca
reaction occurs in the crust of a magnetar, the assumption of
$T\sim 10^{9}$ K is contradictory to the observational values
of $T_{\rm BB}$ in Table 1.  If the direct Urca reaction occurs in
the interior of a magnetar where electrons are relativistic,
neutrons and protons are non-relativistic, the so-called
`standard magnitude' of $Q_{\nu}$ will be higher than $10^{56}$
erg~s$^{-1}$ at least according to our model (see $\S$ 3.2).
By analyzing, we can make a comparison between the observed
values ($Q_{\nu}$ and $L_{\rm X}$) and the conventional
values ($Q_{\nu}$ and $L_{\rm X}$): For a magnetar, a typical
observed soft X-ray luminosity is about 3 $\times 10^{35}$
erg~s$^{-1}$ \citep{tho02}, the mount of total electromagnetic
radiation is about one thirtieth of that of the neutrino
emissivity due to the fact that photons are partly absorbed
by star internal matter,\citep{hey98, tho96, tho02}, then
we obtain the neutrino emissivity $\sim 10^{37}$ erg~s$^{-1}$;
if all emissions of a magnetar are from the modified Urca
reaction (the magnitude of which is about 6-orders lower
than that of the direct Urca reaction), then the calculated
value of $Q_{\nu}$ is $\sim 10^{50}$ erg~s$^{-1}$ or above,
which is far larger than the observed values of $Q_{\nu}$($\sim
10^{37}$ erg~s$^{-1}$), and the calculated value of $L_{X}$ in
the modified Urca process is about $\sim 10^{48}$ erg~s$^{-1}$
or above, accordingly, which is far higher than the observed
value of $L_{\rm X}$, where we assume that the energies of
all particles participating in the process of electron capture
are calculated according to our model. Our model will be a
challenge for the conventional way of calculating neutrino
emissivity, if our analysis is reasonable.

Seeing from Fig. 3 and Table 1, it's obvious that the soft X-ray
luminosity $L_{\rm X}$ of a magnetar increases with increasing $B$.
However, due to the existences of calculation error and observation
error (for example, we ignore the X-ray loss due to being absorbed
by star internal matter), there are still differences between
calculations and observations.  How to improve our model to make
calculations agree well with observations will be considered in
our future work. Despite these previous researches, existing analysis
is approximate and many assumptions are invoked consequently, because
our computational methods are completely new approaches, that need to
be validated empirically.

\section{The calculations of reaction time }
When the electron Fermi energy is much greater than the
threshold energy of inverse $\beta-$decay, the electron
capture reaction will occur immediately.  Thus the ${}^
3P_2$ Cooper pairs will be destroyed quickly by the
outgoing high-energy neutrons, and the anisotropic
superfluid and the superhigh induced magnetic field will
disappear. Employing Eqs.(3-4) can allow us to write a
differential equation
\begin{eqnarray}
&&\frac{d\Gamma}{dt}=\Lambda~S(E_{\rm F}(\rm e)-Q)^{2}E_{\rm F}(\rm e)(E^{2}_{\rm F}(\rm e)\nonumber\\
&&-m^{2}_{\rm e}c^{4})^{\frac{1}{2}}40 \times \frac{1}{4}B^{-\frac{3}{4}}B^
{-\frac{1}{4}}_{\rm cr}\frac{dB}{dt},
\end{eqnarray}
where the approximation $S \approx 1$ is used.  Using the binomial
expansion theorem, the term $(E^{2}_{\rm F}(e)-m^{2}_{\rm e}c^{4})
^{\frac{1}{2}}$ can be expanded as:
\begin{eqnarray}
&&(E^{2}_{\rm F}(\rm e)- m^{2}_{\rm e}c^{4})^{\frac{1}{2}}= E_{\rm F}(e)\times (1- m^{2}_{\rm e}c^{4}/2E^{2}_{\rm F}(\rm e) \nonumber\\
&&-m^{4}_{\rm e}c^{8}/8E^{4}_{\rm F}(e) +  \cdots )\approx 40(B/B_{\rm cr})^{\frac{1}{4}}\nonumber\\
&&\times(1- 542B^{-\frac{1}{2}}- 146932 B^{-1} + \cdots).
\end{eqnarray}
Since 542$B^{-\frac{1}{2}}\sim 10^{-5}$ and 146932$B^{-1}
\sim 10^{-10}$, reserving the first two terms in the bracket
of the expansion above gives
\begin{eqnarray}
&&\frac{d\Gamma}{dt}= 16000\Lambda(1600B^{\frac{1}{4}}
B^{-\frac{5}{4}}_{\rm cr}-4800B_{\rm cr}^{-1}\nonumber\\
&&+3600B^{-\frac{1}{4}}B^{-\frac{3}{4}}_{cr})
(1-542B^{-\frac{1}{2}})\frac{dB}{dt}.
\end{eqnarray}
A neutron star can be treated as a system of magnetic
dipoles $\mu = B R^{3}/2$, where $\mu, B,$ and $R$ are
the polar magnetic moment, the polar magnetic field
strength and the radius of the neutron star, respectively
\citep{sha83}.  If one outgoing neutron in the process of
$e^-+ p\rightarrow n+\nu_e$ is assumed to destroy one
${}^3P_{2}$ Cooper pair, then the decay rate of the
superhigh magnetic field is
\begin{equation}
\frac{dB}{dt} =\frac{2}{R^{3}}\frac{d}{dt}(-q\Gamma 2\mu_{\rm n} n_{\rm e}V
({}^3P_{2}))= \frac{-4\mu_{\rm n} n_{\rm e}V({}^3P_{2})q\Gamma}{R^{3}},
\end{equation}
where $\mu_{\rm n}$= 0.966 $\times 10^{-23}$ erg~G$^{-1}$
is the absolute value of the neutron abnormal magnetic
moment.  Combining  Eq.(10) with Eq.(11) and eliminating
$\Gamma$, we get a second-order differential equation
\begin{eqnarray}
&&\frac{d^{2}B}{dt^{2}}+ \frac{64000\Lambda \mu_{\rm n}
n_{\rm e}V({}^3P_{2})q}{R^{3}}(1600B^{\frac{1}{4}}
B^{-\frac{5}{4}}_{\rm cr}- 4800\nonumber\\
&&B_{cr}^{-1}+ 3600B^{-\frac{1}{4}}B^{-\frac{3}{4}}_{\rm cr})
(1- 542B^{-\frac{1}{2}})\frac{dB}{dt}=0,
\end{eqnarray}
In order to obtain the total electron capture time $t$, this
second-order differential equation can be treated as follows:
firstly, decrease the order of Eq.(12); secondly, apply the
initial reaction condition $B$=2.2346 $\times 10^{14}$ G,
$dB/dt$= 0 to determine the constant of integration $\sim$
1619.4; thirdly, integrating over $B$ gives the expression
for $t$; and lastly, using the integral transform $B^{\frac{1}
{4}}\rightarrow x$ and $dB \rightarrow 4x^{3}dx$ gives the
integral equation

\begin{eqnarray}
&&t=\frac{-R^{3}}{64000 q\Lambda \mu_{\rm n} n_{\rm e}V({}^3P_{2})}
\int_{B^{0.25}_{1}}^{B^{0.25}_{2}}(\frac{1600 \times \frac{4}{5}
x^{5}}{(4.414 \times 10^{13})^{\frac{5}{4}}}\nonumber\\
&&-\frac{4800x^{4}}{4.414 \times 10^{13}}+\frac{3600 \times \frac{4}{3}x^{3}}{(4.414 \times 10^{13})^{\frac{3}{4}}}\nonumber\\
&&-\frac{542 \times 1600 \times \frac{4}{3}x^{3}}{(4.414 \times 10^{13})^{\frac{5}{4}}}+ \frac{542 \times 9600x^{2}}{4.414 \times 10^{13}}-\nonumber\\
&&\frac{542 \times 14400 \times \frac{4}{3}x}{(4.414 \times 10^{13})
^{\frac{3}{4}}}+ 1619.4)^{-1}4x^{3}dx,
\end{eqnarray}
Simplifying Eq.(13) further, we get
\begin{eqnarray}
&&t = \frac{1}{q}\int_{B^{0.25}_{2}}^{B^{0.25}_{1}}(0.502516x^{5}- 4857.23x^{4}+1.25198\nonumber\\
 &&\times10^{7}x^{3}- 453.939x^{3}+ 5.26524 \times 10^{6}x^{2}- 2.03578\nonumber\\
 &&\times 10^{10}x- 6.71187 \times 10^{18})^{-1}4x^{3}dx,
\end{eqnarray}
where $B_{2} > B_{1}$ and the mean value of $q \approx$ 1.94
$\times 10^{-18}$.  Solving Eq.(14) gives $t \approx$ 2.36
$\times 10^{14}$ s = 7.48 $\times 10^{6}$ yrs when $B_{2}$ =
3 $\times 10^{15}$ G and $B_{1}$=2.2346 $\times 10^{14}$ G;
similarly, if $B_{2}$= 3.0 $\times 10^{15}$ G and $B_{1}$=
3.7 $\times 10^{14}$ G, then $t \approx$ 2.66 $\times 10^{14}$
s $\approx$ 7.1 $\times 10^{6}$ yrs corresponding to $L_{\rm X}
\sim$(9.0 $\times 10^{36}$-1.0 $\times 10^{34})$ erg~
s$^{-1}$; if $B_{2}$= 3.7 $\times 10^{14}$ G and $B_{1}$
= 2.25 $\times 10^{14}$ G, then $t \approx$ 3.7 $\times 10^{5}$
yrs corresponding to $L_{\rm X}\sim$ (1.0 $\times 10^{34}$-
1.75 $\times 10^{28}$) erg~s$^{-1}$.

It may be questioned whether there are sufficient anisotropic
${}^3P_{2}$ superfluid Cooper pairs able to be destroyed to
significantly decrease the induced magnetic field in the
magnetar interior.  Allow us to make following calculations:
a neutron star's mass is $\sim$1.4 $M_{\odot}, M_{\odot}
\approx$ 2.0$\times 10^{33}$ g and the mass fraction of
${}^3P_{2}$ anisotropic neutron superfluid in the mass of
neutron star is more than 0.1.  Therefore, $N_{\rm Cooper}\approx$
0.1 $\times$ 1.4 $M_{\odot}/m_{({}^3P_{2})}$=0.14 $\times
M_{\odot}/2m_{\rm n}$=2.8 $\times 10^{32}/(2\times$1.67$\times
10^{-24})\approx$ 8.38 $\times 10^{55}$.  Meanwhile $\Delta
N_{\rm Cooper}$, the total number of ${}^3P_{2}$ Cooper pairs
destroyed by outgoing neutrons is determined by $\Delta
N_{\rm Cooper}= n_{\rm p}V({}^3P_{2})\Gamma_{\rm tot}$, where
$\Gamma_{\rm tot}$, the average total electron number
captured by one proton, can be expressed as
\begin{equation}
 \Gamma_{\rm tot}=\int q \Gamma dt.
\end{equation}
Using Eq.(15), we find
\begin{equation}
 \Gamma_{\rm tot}=\int _{B_{2}}^{B_{1}}\frac{R^{3}}{4\mu_{\rm n}
 n_{\rm e}V({}^3P_{2})}dB \approx 17926,
\end{equation}
where $B_{2}$=3.0 $\times 10^{15}$ G and $B_{1}$=2.2346 $\times
10^{14}$ G.  Thus $\Delta N_{\rm Cooper}$ can be evaluated to be 1.2
$\times 10^{55}$. Be note, in the interior of a NS, the processes
of electron capture and $\beta-$decay exist at the same time
required by electric neutrality, the depleted protons (electrons)
are recycled for many times, so the variable $\Delta Y_{\rm e}$ could
be very small. From evaluations above, it is obvious that $\Delta
N_{\rm Cooper}< N_{\rm Cooper}$, which demonstrates that our results are
reasonable.

\section{ Discussions and Conclusions}
In this paper, we review briefly the evolution of magnetic fields
of magnetars in currently existing magnetar models, investigate
the relation between magnetar soft X-ray Luminosity $L_{\rm X}$ and
magnetic field strength $B$, and calculate magnetar magnetic
field evolution timescale $t$.  From the analysis and the
calculations above, the main conclusions are as follows:

1. Superhigh magnetic fields genuinely produce a remarkable
magnetic broadening of the direct Urca process threshold. The
${}^3P_2$ Cooper pairs will be destroyed quickly by the high-energy
outgoing neutrons via the process of electron capture, so the
induced magnetic field will disappear.

2. The kinetic energy of the outgoing neutrons will be first
transformed into thermal energy and then into radiation energy as
X-rays and soft $\gamma-$rays. The observed X-ray luminosity
$L_{\rm X}$ is related to $\langle E_{\rm n}\rangle$ and $\Gamma_{\rm eff}$,
but is ultimately determined by $B$.

3. The relationship between $B$ and $t$ can be expressed as a
second-order differential equation. If $L_{\rm X0}$ is assumed to be
(1 $\sim$ 9)$\times 10^{36}$ erg~s$^{-1}$, then $q$ is (0.216
$\sim$1.94)$\times 10^{-18}$ and $t$ is (6.7 $\sim$ 0.748)
$\times10^{7}$ yrs, respectively.  The origin of the magnetic
fields in magnetars shows that magnetars could exist, but that
they would be unstable due to the high electron Fermi energy.

Finally, we are hopeful that our calculations can soon be combined
with astrophysical studies of magnetars and new observations, to
provide a deeper understanding of the nature of the superhigh
magnetic fields in magnetars.

\begin{acknowledgments}
We are very grateful to Prof. Qiu-He Peng, Prof. Zi-Gao Dai
and Prof. Yong-Feng Huang for their help of improving this
paper.  This work is supported by Xinjiang Natural Science
Foundation No. 2009211B35, the key Directional Project of
CAS and NNSFC under the project No. 10173020, No. 0673021 and                                                                                                              Chinese National Science Foundation through grant No. 10573005.
\end{acknowledgments}

\appendix
\textbf{Appendix}
\section{ The effect of a superhigh magnetic field on $Y_{\rm p}$  }
As we know, in the case of field-free, for reactions $e^{-}
+p \rightarrow n+ \nu_{\rm e}$ and $n \rightarrow p +e^{-}+\nu^
{-}_{e}$ to take place, there exists the following inequality
among the Fermi momenta of the proton($P_{\rm F}$), the electron
($K_{\rm F}$)and the neutron ($Q_{F}$): $P_{\rm F}+ K_{\rm F}\geq Q_{\rm F}$
required by momentum conservation near the Fermi surface,
$Y_{\rm p}$ is the proton fraction, defined as the mean proton
number per baryon.  Together with the charge neutrality
condition, the above inequality brings about the threshold
for proton concentration $Y_{\rm p}\geq $1/9, this means that,
in the field-free case, direct Urca reactions are strongly
suppressed by Pauli blocking in a system composed of neutrons,
protons, and electrons ($n-e-p$ system).  In the core of a NS,
where $\rho \geq 10^{15}$ g~cm$^{-1}$ and $Y_{\rm p}$ could be
higher than 0.11, direct Urca processes could take place
\citep{bai99, lai91, yak01}.  However, when in a superhigh
magnetic field $B\gg B_{\rm cr}$, things could be quite different
if we take into account of the effect of the intense
magnetic field on $Y_{\rm p}$.  The effect of a superhigh
magnetic field on the NS profiles is gained by applying
equations of state (EOS) to solve the Tolman-Oppenheimer-Volkoff
equation \citep{sha83}.  In the paper of Chakrabarty S. et al
(1997), employing a relativistic Hartree theory, authors
investigated the gross properties of cold symmetric matter
and matter in $\beta-$equilibrium under the influences of
ultrastrong magnetic fields. The main conclusions of the paper
include: There could be an extremely intense magnetic field $\sim
10^{20}$ G inside a NS; $Y_{\rm p}$ is a strong function of $B$ and
$\rho$; when $B$ is near to $B_{\rm cr}^{p}$, the value of $Y_{\rm p}$
is expected to be considerably enhanced, where $B_{\rm cr}^{p}$ is
the quantum critical magnetic field of protons ($\sim$1.48
$\times 10^{20}$ G); by strongly modifying the phase spaces of
protons and electrons, magnetic fields of such magnitude ($\sim
10^{20}$ G) can cause a substantial $n \rightarrow p$ conversion,
as a result, the system is converted to highly proton-rich matter
with distinctively softer EOS, compared to the field-free case.
Though magnetic fields of such magnitude inside NSs are
unauthentic, and are not consistent with our model ($B\sim
10^{14-15}$ G), their calculations are useful in supporting
our following assumption: when $B\sim 10^{14-15}$ G, the value
of $Y_{\rm p}$ may be enhanced, and could be  slightly higher than
the mean value of $Y_{\rm p}$ inside a NS ($\sim$ 0.05).  Based
on this assumption, we can gain a concise expression for $E_{\rm F}(\rm e)$,
$E_{\rm F}(\rm e)$=39.9$(B/B_{\rm cr})^{\frac{1}{4}}(\frac{\rho}{\rho_{0}}
\frac{Y_{\rm p}}{0.085})^{\frac{1}{4}}$ MeV $\simeq $40$(B/B_{\rm cr})
^{\frac{1}{4}}(\frac{\rho}{\rho_{0}}\frac{Y_{\rm p}}{0.085})
^{\frac{1}{4}}$ MeV by solving equation(19) of Gao Z. F et al (2010b).

\section{A wrong conclusion on the Fermi energy of electrons
 in superhigh magnetic fields}

It is universally accepted that $E_{\rm F}(\rm e)$ decreases with
increase in $B$ in the case of superhigh magnetic fields.
The reasons for this are as follows: when in the presence of
a uniform external magnetic field along the $z-$axis,
solving the non-relativistic Schr\"{o}dinger Equation
for electrons gives electron energy level
\begin{equation}
  E_{\rm e}(p_{z}, B, n, \sigma)= p_{z}^{2}c^{2}/2m_{\rm e}
 +(2n+ 1+ \sigma)\hbar\omega_{B},
 \end{equation}
where $\hbar\omega_{B}= 2\mu_{\rm e}B$, $\omega_{B}= eB/m_{\rm e}c$
is the well-known non-relativistic electron cyclotron frequency,
(c.f Page 460 of Quantum Mechanics \citep{lan65}). In the interval
[$p_{z}$ $p_{z}+ dp_{z}$] along the magnetic field, for a non-relativistic
electron gas, the possible microstate numbers are calculated
by $N_{\rm pha}(p_{z})=\frac{eB}{4\pi \hbar^{2}}\frac{dp_{z}}{c}$.  Therefore,
we obtain
\begin{equation}
N_{\rm pha}= \int_{0}^{p_{\rm F}}N_{\rm pha}(p_{z})dz = \frac{eB}{4\pi
\hbar^{2}}\frac{E_{\rm F}(\rm e)}{c^2},
\end{equation}
where the solution of Eq.(B1) is used (also c.f. Page 460 of
Quantum Mechanics \citep{lan65}).  In the light of the Pauli
exclusion principle, the electron number density should be equal
to the its microstate density, one can gain the expression
\begin{equation}
N_{\rm pha}= n_{\rm e}= \frac{eB}{4\pi \hbar^{2}}\frac{E_{\rm F}(e)}{c^2}=
N_{A}\rho Y_{\rm e}.
 \end{equation}
From Eq.(B3), it is easy to see $E_{\rm F}(\rm e) \propto B^{-1}$
when $n_{\rm e}$ is given.  We then ask why such a phenomenon
exists.  After careful analysis, we find that the solution of
the non-relativistic electron cyclotron motion equation
$\hbar\omega_{B}$ is wrongly (or unsuitably) applied to
calculate energy state density in a relativistic degenerate
electron gas.  It's interesting to note that, in the Page 12
of Canuto \& Chiu (1971), in order to evaluate the degeneracy
of the $n$-th Landau level $\omega_{n}$, they firstly introduce
the cylindrical coordinates ($p_{\perp},\phi$) where $\phi
=\arctan p_{x}/p_{y}$, then gain an approximate relation
\begin{equation}
\omega_{n}=(2\pi\hbar)\int_{0}^{2\pi}d\phi\int_{A< p_{\perp}^{2}< B}p_{\perp}dp_{\perp}
= 2\pi(2\pi\hbar)^{-2}\frac{1}{2}(B-A)= \frac{1}{2\pi}(\hbar/m_{\rm e}c)^{-2}\cdot(B/B_{\rm cr}),
\end{equation}
where $A= m^{2}c^{2}\frac{B}{B_{\rm cr}}2n$ and $B= m^{2}c^{2}
\frac{B}{B_{\rm cr}}2(n+ 1)$\citep{can71}. In the paper, authors
stated clearly that this relation is hold only when $B$ =0, in
other words, this relation is just an approximation in the case
of weak magnetic field ($B\ll B_{\rm cr}$).  Surprisingly, this
relation has been misused for nearly 40 years since then. Even
in some textbooks on statistical physics, the statistical weight
is calculated unanimously using the expression
\begin{equation}
 \frac{1}{h^2}\int dp_{x}dp_{y}= \frac{1}{h^2}\pi p_{\perp}^{2}
 \mid _{n}^{n+1}= \frac{4\pi m_{\rm e}\mu_{\rm e}B}{h^{2}}.
 \end{equation}
This expression will also cause the wrong deduction: $E_{\rm F}(\rm e)\propto
B^{-1}$, which is exactly the same as that from Eq.(B3). The
sources of the wrong deduction lie in :  no consideration of the
conditions of Eq.(B4); the assumption that the torus located between
the $n$-th Landau level and the $(n+1)$-th Landau level in momentum
space is ascribed to the $(n+1)$-th Landau level.  For the latter, the
electron energy (or momentum) will change continuously in the direction
perpendicular to the magnetic field, which is contradictory to the
quantization of energy (or momentum) in the presence of superhigh
magnetic field.  Actually, electrons are relativistic and degenerate
in the interior of a NS, so that Eq.(B1) and Eq.(B2) are no longer
applicable, and need to be modified.  Considering this, we replace
Eq.(B1) and Eq.(B2) by Eq.(B6) and Eq.(B7), respectively:
\begin{equation}
  E_{\rm e}^{2}(p_{z}, B, n, \sigma)= m_{\rm e}^{2}c^{4}+ p_{z}^{2}c^{2}
 +(2n+ 1+ \sigma)2m_{\rm e}c^{2}\mu_{\rm e}B,
  \end{equation}
\begin{equation}
N_{\rm pha}= \frac{2\pi}{h^{3}}\int dp_{z}\sum_{n=0}^{n_{\rm m}(p_z, \sigma, B^{*})}\sum g_{n}
 \int \delta(\frac{p_{\perp}}{m_{\rm e}c}-[(2n+ 1+ \sigma)B^{*}]^{\frac{1}{2}})p_{\perp}dp_{\perp},
 \end{equation}
where $\delta(\frac{p_{\perp}}{m_{\rm e}c}-[(2n+ 1+ \sigma)B^{*} ]^
{\frac{1}{2}})$ is the Dirac $\delta$-function.  Failing to do
so would cause us to reach the wrong conclusion: $E_{\rm F}(\rm e)$
decreases with the increase in $B$ when in an intense field.
(Cited partly from Gao Z. F et al 2010b.)

\end{document}